\documentclass[twocolumn,aps,prb,showpacs,preprintnumbers,amsmath,amssymb, superscriptaddress]{revtex4-1}
\usepackage{bm}
\usepackage{graphicx}
\usepackage{color,xspace}
\usepackage{ulem}
\usepackage{float}

\newenvironment{addendum}{%
   \setlength{\parindent}{0in}%
   \small%
   \begin{list}{Acknowledgements}{%
       \setlength{\leftmargin}{0in}%
       \setlength{\listparindent}{0in}%
       \setlength{\labelsep}{0em}%
       \setlength{\labelwidth}{0in}%
       \setlength{\itemsep}{12pt}%
       }
   }
   {\end{list}\normalsize}

\begin{document}

\title{Direct observation of multiple topological phases in the iron-based superconductor Li(Fe,Co)As}

\author{Peng~Zhang}
\affiliation{Institute for Solid State Physics, University of Tokyo, Kashiwa, Chiba 277-8581, Japan}

\author{Xianxin Wu}
\affiliation{Theoretical Physics, University of W\"urzburg Am Hubland 97074 W\"urzburg Germany }

\author{Koichiro Yaji}
\affiliation{Institute for Solid State Physics, University of Tokyo, Kashiwa, Chiba 277-8581, Japan}

\author{Guangyang Dai}
\affiliation{Beijing National Laboratory for Condensed Matter Physics and Institute of Physics, Chinese Academy of Sciences, Beijing 100190, China}

\author{Xiancheng Wang}
\affiliation{Beijing National Laboratory for Condensed Matter Physics and Institute of Physics, Chinese Academy of Sciences, Beijing 100190, China}

\author{Changqing Jin}
\affiliation{Beijing National Laboratory for Condensed Matter Physics and Institute of Physics, Chinese Academy of Sciences, Beijing 100190, China}

\author{Jiangping Hu}
\affiliation{Beijing National Laboratory for Condensed Matter Physics and Institute of Physics, Chinese Academy of Sciences, Beijing 100190, China}

\author{Ronny Thomale}
\affiliation{Theoretical Physics, University of W\"urzburg Am Hubland 97074 W\"urzburg Germany }

\author{Takeshi Kondo}
\affiliation{Institute for Solid State Physics, University of Tokyo, Kashiwa, Chiba 277-8581, Japan}

\author{Shik Shin}
\affiliation{Institute for Solid State Physics, University of Tokyo, Kashiwa, Chiba 277-8581, Japan}

\date{\today}

\maketitle

\textbf{ Topological insulators/semimetals and unconventional iron-based superconductors have attracted major attentions in condensed matter physics in the past 10 years\cite{HasanRMP2010, QiRMP2011, BansilRMP2016, HosonoJACS2008, JohnstonAIP2010, StewartRMP2011}. However, there is little overlap between these two fields, although the combination of topological states and superconducting states will produce more exotic topologically superconducting states and Majorana bound states (MBS), a promising candidate for realizing topological quantum computations. With the progress in laser-based spin-resolved and angle-resolved photoemission spectroscopy (ARPES) with very high energy- and momentum-resolution, we directly resolved the topological insulator (TI) phase and topological Dirac semimetal (TDS) phase near Fermi level ($E_F$) in the iron-based superconductor Li(Fe,Co)As. The TI and TDS phases can be separately tuned to $E_F$ by Co doping, allowing a detailed study of different superconducting topological states in the same material. Together with the topological states in Fe(Te,Se)\cite{TSC, TDS}, our study shows the ubiquitous coexistence of superconductivity and multiple topological phases in iron-based superconductors, and opens a new age for the study of high-Tc iron-based superconductors and topological superconductivity. }

High-Tc iron-based superconductors have multiple bands near $E_F$, which increases the difficulty for the understanding of the unconventional pairing. On the other side, it also produces rich electronic phases, of which a recent example is the TI phase at $E_F$ in the iron-based superconductor Fe(Te,Se)~\cite{TSC}, which provides a simple solution to realize 2D topological superconductivity and MBS, rather than the complex low-$T_c$ heterostructures \cite{FuPRL2008, MourikScience2012, YazdaniScience2014, AlbrechtNature2016}. It is predicted that this topological state will produce a phase transition between topological superconductivity and trivial superconductivity with $E_F$ shifting \cite{XuPRL2016}, which could be useful to optimize the topological superconductivity. In addition, another TDS phase slightly above $E_F$ has also been unveiled by spin-resolved ARPES (SARPES) and magnetoresistance measurements \cite{TDS}.
Further studies on the above topics would require to dope the material, which is difficult for Fe(Te,Se). In this paper, we focus on the iron-based superconductor LiFe$_{1-x}$Co$_x$As, whose Fermi level can be easily shifted by Co doping without affecting the quasiparticle coherence \cite{PitcherJACS2010, MiaoPRB2014, MiaoNC2015}. Similar to that of Fe(Te,Se), a TI phase near $E_F$ also exists, consistent with the first-principles calculations. Furthermore, we directly observed the TDS phase with $x = 9\%$ sample in the ARPES measurement. Our study proves the universality of topological states in iron-based superconductors and  provides a simple platform to realize tunable and possibly multiple topological superconducting states.

\begin{figure*}[!htb]
\begin{center}
\includegraphics[width=0.9\textwidth]{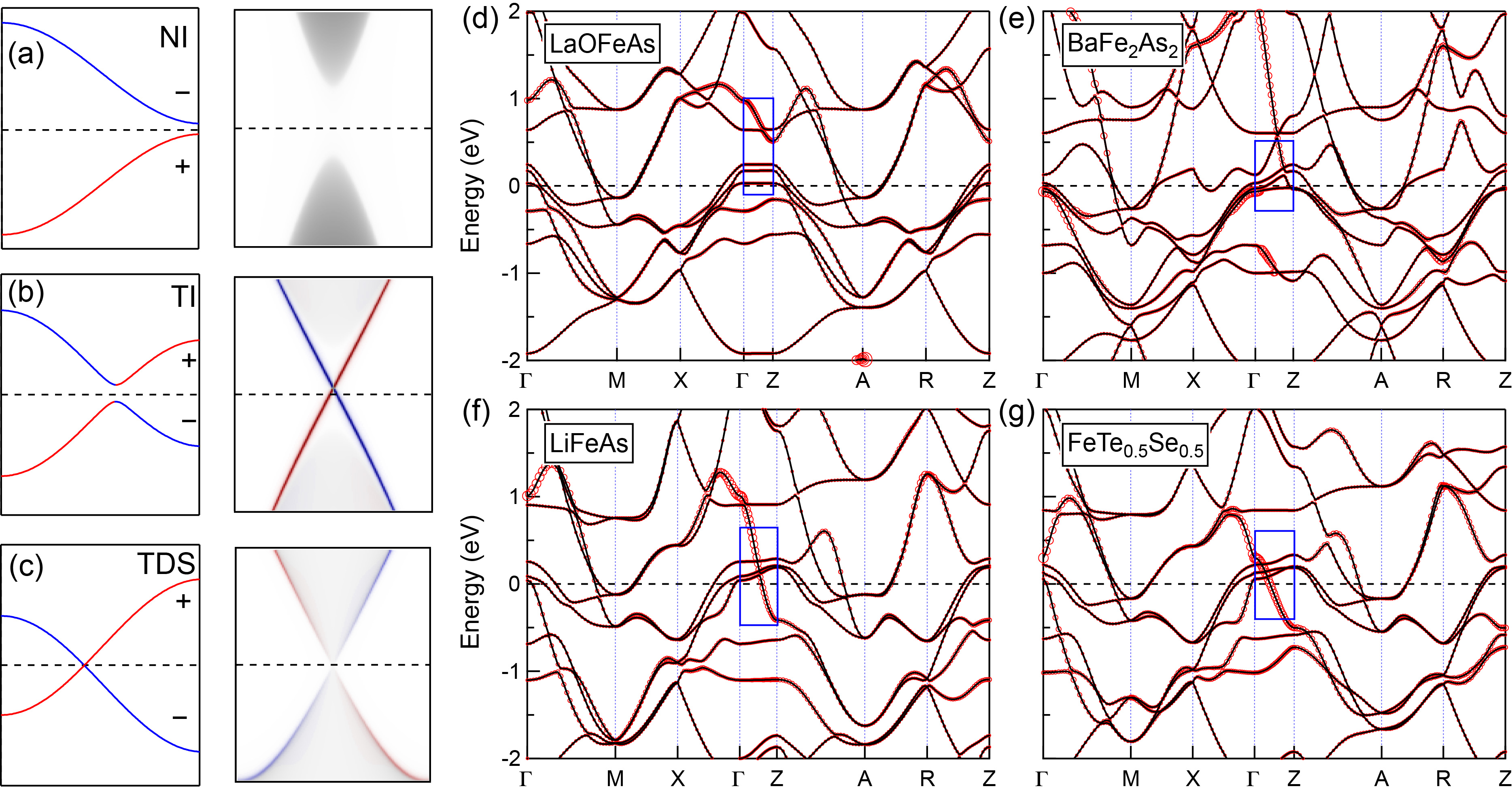}
\end{center}
 \caption{\label{band}  \textbf{Different topological phases and band structures of iron-based superconductors.} (a - c) The out-of-plane dispersion and the corresponding surface spectrum, for NI (a), TI (b) and TDS (c). We overlapped the spin-polarized surface spectrum (blue and red) on top of the surface spectrum (grey). (d - g) Band structures of LaOFeAs, BaFe$_2$As$_2$, LiFeAs and FeTe$_{0.5}$Se$_{0.5}$, respectively. No band inversion between $p_z$ and $d_{xz}$/$d_{yz}$ is found in LaOFeAs, while there are band inversions in BaFe$_2$As$_2$, LiFeAs and FeTe$_{0.5}$Se$_{0.5}$.}
\end{figure*}

Normal insulator (NI), TI and TDS are three different topological phases. From the view of the band structure, if there is no band inversion in whole BZ, the material will be a normal insulator, with no spin-polarized surface states, as displayed in Fig. 1(a). In a simple case, if there is a single band inversions in the whole BZ, the material will be topologically non-trivial, with spin-polarized Dirac-cone band. In a TI, the bulk band gap leads to well-defined surface states, with explicit spin helicity [Fig. 1(b)]. Instead, in a TDS,  the band crossing is protected by the crystal symmetry, and the surface Dirac  band generally overlap with the bulk Dirac band on the (001) surface \cite{WangPRB2012, WangPRB2013}. The spin-polarization magnitude of the (001) surface states generally show a gradual increase with the increased distance from the Dirac point, as shown in Fig. 1(c). The three phases may coexist and appear in one material.

In the previous papers, we show that there are band inversions between $p_z$ and $d_{xz}$/$d_{yz}$ bands in Fe(Te,Se), resulting in TI and TDS phases \cite{WangPRB2015, XuPRL2016, TSC, TDS}. Here we further check the band structure of the four major classes of iron-based superconductors, focusing on the band inversions of $p_z$ and $d_{xz}$/$d_{yz}$ bands along $k_z$.
 The separations between the adjacent FeAs/FeSe layers $\Delta_d$ are 8.741, 6.508, 6.364 and 5.955 for LaOFeAs (1111), BaFe$_2$As$_2$ (122), LiFeAs (111) and FeTe$_{0.5}$Se$_{0.5}$ (11), respectively \cite{JohnstonAIP2010, WangPRB2015}.
This separation directly determine the interlayer $pp$ coupling, which will affect the band width of the $p_z$ band along $\Gamma$Z \cite{WangPRB2015}. The parameter $\Delta_d$ for LaOFeAs is much larger than and other three systems BaFe$_2$As$_2$, LiFeAs and Fe(Te,Se), resulting in a small $p_z$ dispersion for LaOFeAs but large $p_z$ dispersions for BaFe$_2$As$_2$, LiFeAs and Fe(Te,Se). In Fig.1(d - g), we display the band structures of the four classes, where the size of red circles represent the weights of As/Se $p_z$ orbital. It is clear that along the $\Gamma$Z direction there is no band inversion for LaOFeAs, but there are band inversions for BaFe$_2$As$_2$, LiFeAs and Fe(Te,Se). However, the existence of band inversions should be checked by experiments. In quasi-two dimensional family and 2D thin films, such as Fe(Te,Se) film, band inversion can happen at both $\Gamma$ and Z points. The inplane lattice parameter $a$ (or Se/Te height) affects the intralayer $pd$ coupling and determine the position of $p_z$ band at $\Gamma$. By reducing the parameter $a$ (increasing the Se/Te height), $p_z$ band can sink below $d_{xz/yz}$ band at $\Gamma$, generating a band inversion and realizing a 2D topological insulator phase \cite{WuPRB2016, DingSB2017}

\begin{figure*}[!htbp]
\begin{center}
\includegraphics[width=0.85\textwidth]{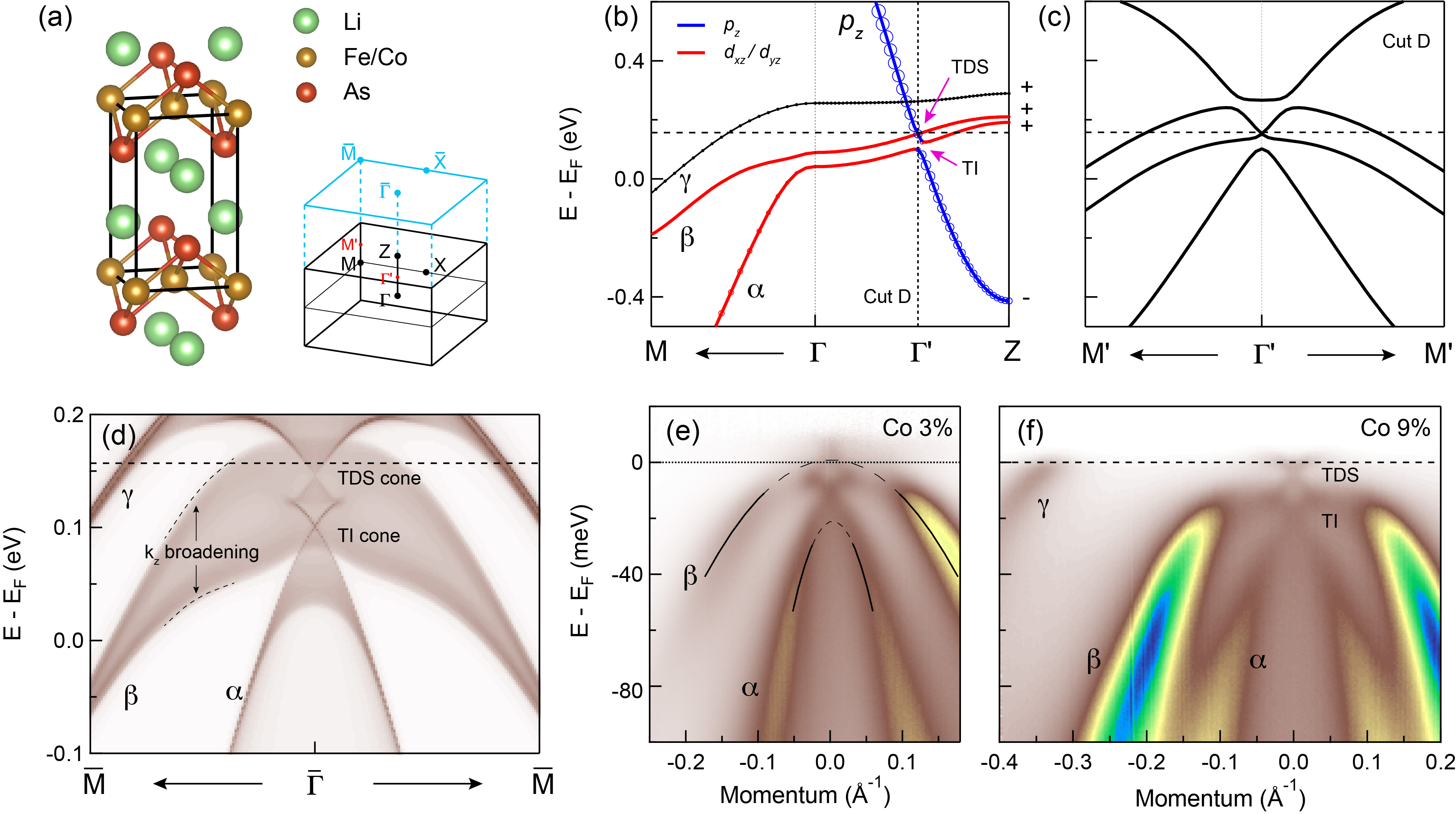}
\end{center}
 \caption{\label{spin} \textbf{Electronic structure of Li(Fe,Co)As. } (a) Left: Crystal structure of Li(Fe,Co)As.  Right: High symmetry points in the 3D BZ and (001) surface BZ. (b) Zoom-in view of the LiFeAs band dispersion along $\Gamma$M and $\Gamma$Z. The red color marks the $d_{xz}$/$d_{yz}$ bands, and the blue color marks the $p_z$ band. The marker size relates to the weight of the $p_z$ orbital character. The real crossing between $\beta$ band and $p_z$ band produces a TDS phase, while the voided crossing between $\alpha$ band and $p_z$ produces a TI phase. The crossing between $\gamma$ band and $p_z$ band also produces a TDS phase far above $E_F$. (c) In-plane band structure at Cut D, where the bulk Dirac cone of the TDS phase is shown. (d) (001) surface spectrum of LiFeAs. There is a large broadening in the spectrum at the $\alpha$ and $\beta$ band tops with reduced intensity, due to the $k_z$ dispersion. (e) ARPES intensity plot of LiFe$_{1-x}$Co$_x$As ($x$ = 3\%) at 15 K, with $p$-polarized photons. The spectrum is divided by the corresponding Fermi function. (f) ARPES intensity plot of LiFe$_{1-x}$Co$_x$As ($x$ = 9\%) at 10 K, with $p$-polarized photons. The two Dirac cones are similar to the ones in the calculated surface spectrum in (d). }
\end{figure*}

Compared to BaFe$_2$As$_2$, LiFeAs has no magnetic and structural transitions and it is easy to dope electrons with Co. Thus we focus on the band structure of Li(Fe,Co)As.
Band inversion in LiFeAs and NaFeAs has been resolved in previous papers \cite{WangPRB2015, WatsonPRB2018}. However, the related topological states are still unclear due to the low resolutions. The laser-based high-resolution ARPES and SARPES make it possible to directly resolve the topological states.
The crystal structure is shown in Fig.2(a). Since the Li atoms are quite close to the FeAs plane, the lattice parameter $c$ is comparable to that of Fe(Te,Se). Thus a large $p_z$ dispersion is also present in LiFeAs, as shown in the zoom-in view of the band structure along $k_z$ in Fig.2(b). Since the $p_z$ band has a ``$-$'' parity, while $\alpha$, $\beta$ and $\gamma$ bands all have a ``$+$'' parity, there are multiple band inversions. The real crossing between $\beta$ band and $p_z$ band produces a TDS phase, while the voided crossing between $\alpha$ band and $p_z$ band produces a TI phase. The bulk Dirac cone from TDS phase can be seen from the in-plane band structure at Cut D, as shown in Fig.2(c). The surface Dirac cone from both TI and TDS phases can be seen from the (001) surface spectrum in Fig.2(d).
In the surface spectrum, all dispersions at different $k_z$ of one band will show up. Since bulk bands have $k_z$ dispersions and surface bands do not, the bulk bands generally appears as broad continuums and the surface bands appear as sharp features. In Fig.2(d), The $\alpha$ and $\beta$ band tops and whole $p_z$ band appear as broad and weak continuums due to their $k_z$ dispersions, while the the two surface Dirac cones from TI and TDS are very sharp.

In Fe(Te,Se), only TI phase is directly observed, while the TDS phase is confirmed by some indirect evidences \cite{TSC, TDS}. Instead, here we observed both TDS and TI surface Dirac cones from ARPES experiment on Li(Fe,Co)As. The ARPES band structure of Li(Fe,Co)As with 3\% Co is displayed in Fig.2(e). Despite the surface Dirac cone from the TI phase observed previously in Fe(Te,Se) \cite{TSC}, the second surface Dirac cone from TDS phase, which is above $E_F$, shows up in the ARPES spectrum after dividing the corresponding Fermi function.
We further checked Li(Fe,Co)As sample with 9\% Co, and display in Fig.2(f). As expected, the TDS surface cone shifts down and the full cone clearly shows up, directly confirming the existence of the TDS phase.
The bulk band tops of $\alpha$ and $\beta$ bands are not visible either in Fig.2(e) or Fig.2(f), similar to the calculated surface spectrum in Fig.2(d), which means that the 7-eV laser-ARPES spectrum of Li(Fe,Co)As is better described by the surface spectrum, rather than the bulk band structure at a specific $k_z$.

\begin{figure*}[!htb]
\begin{center}
\includegraphics[width=.9\textwidth]{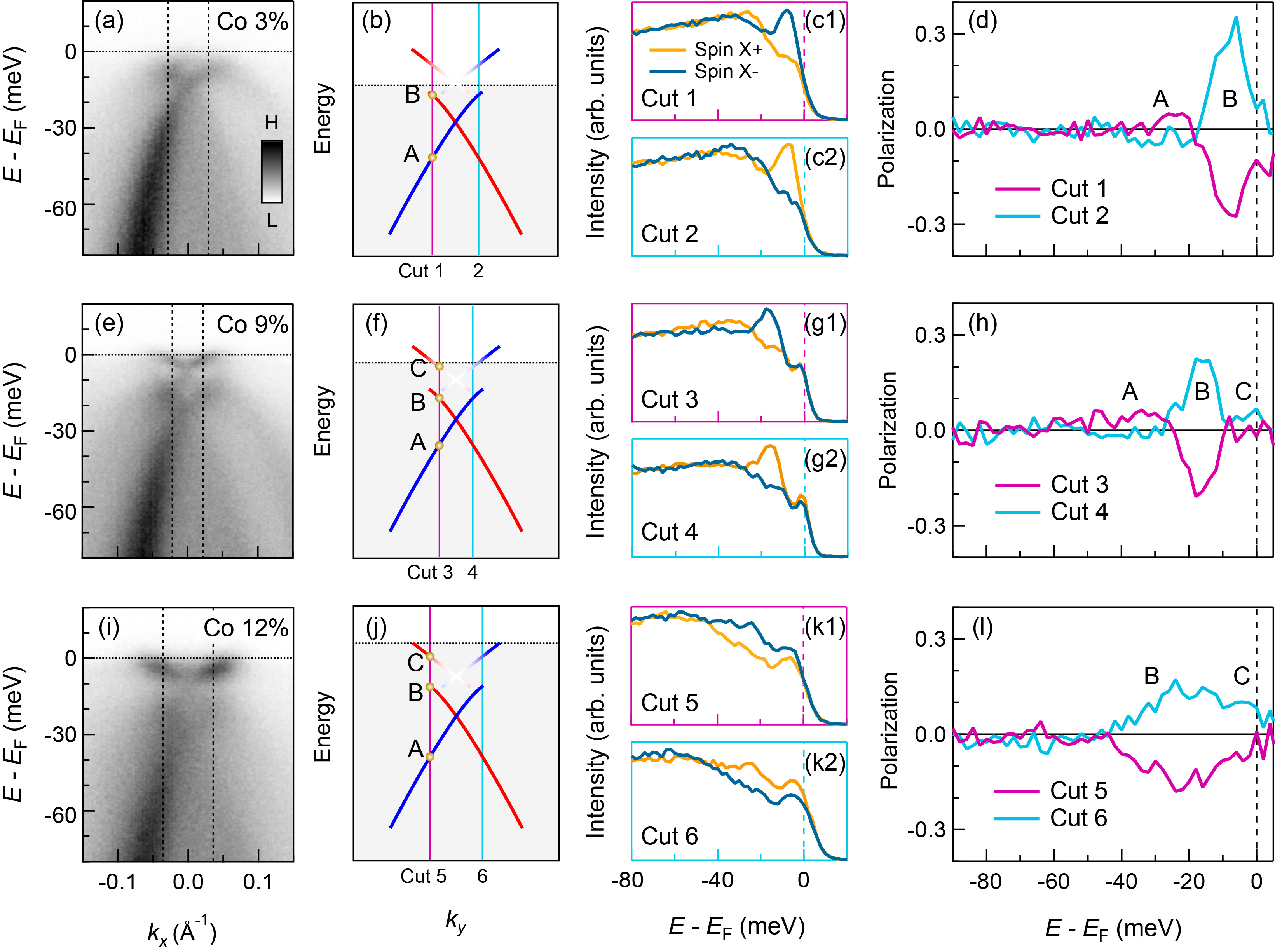}
\end{center}
 \caption{\label{theory} \textbf{Spin polarization of the surface Dirac cones from TI and TDS phases in Li(Fe$_{1-x}$Co$_x$)As.} (a) Intensity plot along $k_x$ for the $x$ = 3\% sample, with $p$-polarized photons. (b) Sketch of the spin-polarized surface bands along $k_y$. The blue and red colors stand for opposite spin polarizations, as confirmed by the data in (c  - d). We measured spin resolved EDCs at the cuts indicated by cyan (Cut 1) and pink (Cut 2) lines. The two lines are also duplicated in (a) as dashed lines. Note that the two cuts in (b) are at different $k_y$, which is to avoid the intensity asymmetry in (a) (Supplementary Information). Because of the $C_4$ symmetry, the band structures along $k_y$ and along $k_x$ are the same, despite the intensity difference. (c) Spin-resolved EDCs at Cut 1 (c1) and Cut 2 (c2), respectively. (d) Spin polarization at Cut 1 and Cut 2. The spin along $x$ direction is measured. A and B correspond to the positions indicated in (b). (e - h) Same as (a - d), but for samples with $x$ = 9\%. (i - l) same as (a - d), but for samples with $x$ = 12\%. }
\end{figure*}

As we show in Fig.1(b,c), both TI and TDS surface Dirac cone should be spin polarized. Thus we can use SARPES~\cite{YajiRSI2016} to double check their surface nature. As shown in Fig.3, we measured three different compositions of LiFe$_{1-x}$Co$_x$As with SARPES. The spin-integrated band structures are the same as the ones in Fig.2, except some intensity difference induced by the experimental geometries (Supplementary Information). A pair of spin-polarized EDCs along $k_y$ was measured, with the EDC positions illustrated in Fig.3(b, f, j).
We first focus on the spin polarization of the TI Dirac cone, which can be obtained from both $x$ = 3\% and $x$ = 9\% samples, as shown in Fig.3(a - d) and (e - h). In both samples, the lower part of the cone (position A) shows opposite spin polarization with the upper part (position B), and the left part of the cone (Cut 1 and Cut 3) shows opposite spin polarization with the right part (Cut 2 and Cut 4), as expected for the spin-polarization from a Dirac cone. The direction of the spin polarization indicates that the Dirac cone has a left-hand helicity, the same as that of Fe(Te,Se) and most TI.
As we discussed in Fig.1, the TDS surface Dirac cone is similar to that of TI, whose spin polarization can be obtained from both $x$ = 9\% and $x$ = 12\% samples, as shown in Fig.3(e - h) and (i - l). We notice that the lower part of the TDS surface cone shows very weak intensity [Fig.3(e, i)], and thus focus on the spin polarization of the upper part (position C). In the $x$ = 9\% sample [Fig.3(g - h)], a weak spin-polarization at position C is resolved. In the $x$ = 12\% sample [Fig.3(k - l)], a pair of EDCs more distant from the conal point is measured and a larger spin polarization is observed. These results are consistent with the model calculations in Fig.1(c): The spin polarization magnitude of the TDS surface cone is smaller near the conal point, while larger far from the conal point. The experimental spin polarizations of the TI and TDS surface cones are summarized in Fig.3 (b, f, j). The spin polarizations confirm that the two Dirac cones indeed come from the surface states.

The multiple topological phases in iron-based superconductors provide a prefect platform for the combined study of topological states and high-$T_c$ superconductivity. In the $x < 3\%$ samples, the topological surface state of TI phase crosses $E_F$, and the surface becomes 2D topologically superconducting when $T < T_c$, due to the $s$-wave pairing in bulk \cite{FuPRL2008}. By changing the Co concentration, it is possible to study the topological superconducting phase transition proposed in Ref. \citenum{XuPRL2016}. 
On the other side, in the $x \ge 9\%$ samples, the Dirac cone of TDS phase crosses $E_F$, which may likely induce a bulk topologically superconducting state, as a consequence of the orbit-momentum locking \cite{SatoPRL2015, SatoPRB2016, TDS}. In Li(Fe,Co)As, there are trivial bulk bands overlapping with the topological states, which thus can couple with the Dirac bands. It is argued that  this coupling can help to stabilize the topological superconductivity over a larger range of doping\cite{HughesPRL2012}. Further theoretical and experimental studies are needed to understand these interactions.

Our findings of the topological states in Li(Fe,Co)As, together with that in Fe(Te,Se), prove the ubiquitous existence of topological phases in iron-based superconductors. The simple structure, multiple topological phases and easily tunable $E_F$ make Li(Fe,Co)As single crystal an ideal platform for the study of topological superconductivity and MBS, and may eventually advance the research of topological quantum computations. 

\textbf{Methods}

High quality single crystals of LiFe$_{1-x}$Co$_x$As were synthesized by self-flux method. All the ARPES measurement are based on 6.994-eV laser. The high resolution ARPES measurements were performed on a spectrometer with a VG-Scienta R4000WAL electron analyzer. The energy resolution of the system was set to $\sim$ 5 meV. The spin-resolved ARPES measurements were carried out with twin very-low-energy- electron-diffraction (VLEED) spin detectors. The energy resolution for the spin-resolved mode was set to $\sim$ 6 meV for $x = 3\%$ and 9\% samples, and $\sim$ 12 meV for $x = 12\%$ sample.

Our Density functional theory calculations employ the projector augmented
wave  method encoded in Vienna ab initio simulation
package\cite{Kresse1993,Kresse1996,Kresse1996prb}, and the local density approximation for the exchange correlation functional is used\cite{Perdew1996prl}.
Throughout this work, the cutoff energy of 500 eV is taken for expanding the wave functions into plane-wave basis. In the
calculation, the Brillouin zone is sampled in the $k$ space within
Monkhorst-Pack scheme\cite{Monkhorst1976}. The number of these $k$ points depends on materials: 11 $\times$ 11 $\times$  5 and 9$\times$ 9 $\times$ 9 for LaOFeAs, LiFeAs, Fe(Te,Se) conventional cells and BaFe$_2$As$_2$ primitive cell, respectively. The spin-orbit coupling was included in the self-consistent calculations of electronic structure.

\begin{addendum}
\item We acknowledge K. Asakawa, A. Harasawa, Y. Hesagawa, K. Ishizaka, N. Mitsuishi, M. Sakano, Y. Yoshida for experimental assistance. This work was supported by the Photon and Quantum Basic Research Coordinated Development Program from MEXT, JSPS (KAKENHI Grant No. 25220707).
The work in W\"urzburg is supported by European Research Council (ERC) Starting Grant TOPOLECTRICS, ERC-StG-Thomale-TOPOLECTRICS-336012.

\item[Competing Interests] The authors declare that they have no competing financial interests.
\item[Correspondence] Correspondence and request for materials should be addressed to P.Z. or S.S. (emails: zhangpeng@issp.u-tokyo.ac.jp, shin@issp.u-tokyo.ac.jp)
\item[Author contributions] P.Z. did the ARPES measurements and analyzed the data with help from K.Y., T.K. and S.S.. X.Wu, J.H. and R.T. did the theory calculations. G.D., X.W. and C.J. synthesized the samples. All authors discussed the manuscript. P.Z. and S.S. supervised the project.
\end{addendum}


\begin{thebibliography}{10}
\expandafter\ifx\csname url\endcsname\relax
  \def\url#1{\texttt{#1}}\fi
\expandafter\ifx\csname urlprefix\endcsname\relax\def\urlprefix{URL }\fi
\providecommand{\bibinfo}[2]{#2}
\providecommand{\eprint}[2][]{\url{#2}}

\bibitem{HasanRMP2010}
\bibinfo{author}{Hasan, M.~Z.} \& \bibinfo{author}{Kane, C.~L.}
\newblock \bibinfo{title}{{Colloquium: Topological insulators}}.
\newblock \textit{\bibinfo{journal}{{Rev. Mod. Phys.}}}
  \textbf{\bibinfo{volume}{{82}}}, \bibinfo{pages}{{3045}}
  (\bibinfo{year}{{2010}}).

\bibitem{QiRMP2011}
\bibinfo{author}{Qi, X.-L.} \& \bibinfo{author}{Zhang, S.-C.}
\newblock \bibinfo{title}{{Topological insulators and superconductors}}.
\newblock \textit{\bibinfo{journal}{{Rev. Mod. Phys.}}}
  \textbf{\bibinfo{volume}{{83}}}, \bibinfo{pages}{{1057}}
  (\bibinfo{year}{{2011}}).

\bibitem{BansilRMP2016}
\bibinfo{author}{Bansil, A.}, \bibinfo{author}{Lin, H.} \&
  \bibinfo{author}{Das, T.}
\newblock \bibinfo{title}{{Colloquium: Topological band theory}}.
\newblock \textit{\bibinfo{journal}{{Rev. Mod. Phys.}}}
  \textbf{\bibinfo{volume}{{88}}}, \bibinfo{pages}{{021004}}
  (\bibinfo{year}{{2016}}).

\bibitem{HosonoJACS2008}
\bibinfo{author}{Kamihara, Y.}, \bibinfo{author}{Watanabe, T.},
  \bibinfo{author}{Hirano, M.} \& \bibinfo{author}{Hosono, H.}
\newblock \bibinfo{title}{{Iron-Based Layered Superconductor
  La[O$_{1-x}$F$_x$]FeAs ($x$ = 0.05-0.12) with $T_c$ = 26 K}}.
\newblock \textit{\bibinfo{journal}{{J. Am. Chem. Soc.}}}
  \textbf{\bibinfo{volume}{{130}}}, \bibinfo{pages}{{3296}}
  (\bibinfo{year}{{2008}}).

\bibitem{JohnstonAIP2010}
\bibinfo{author}{Johnston, D.~C.}
\newblock \bibinfo{title}{{The puzzle of high temperature superconductivity in
  layered iron pnictides and chalcogenides}}.
\newblock \textit{\bibinfo{journal}{{Adv. Phys.}}}
  \textbf{\bibinfo{volume}{{59}}}, \bibinfo{pages}{{803}}
  (\bibinfo{year}{{2010}}).

\bibitem{StewartRMP2011}
\bibinfo{author}{Stewart, G.~R.}
\newblock \bibinfo{title}{{Superconductivity in iron compounds}}.
\newblock \textit{\bibinfo{journal}{{Rev. Mod. Phys.}}}
  \textbf{\bibinfo{volume}{{83}}}, \bibinfo{pages}{{1589}}
  (\bibinfo{year}{{2011}}).

\bibitem{TSC}
\bibinfo{author}{Zhang, P.} \textit{et~al.}
\newblock \bibinfo{title}{{Observation of topological superconductivity on the
  surface of iron-based superconductor}}  (\bibinfo{year}{{2017}}).
\newblock \urlprefix\url{{http://arxiv.org/abs/1706.05163v1}}.
\newblock \eprint{{1706.05163v1}}.

\bibitem{TDS}
\bibinfo{author}{Zhang, P.} \textit{et~al.}
\newblock \bibinfo{title}{{Topological Dirac semimetal phase in the iron-based
  superconductor Fe(Te,Se)}} .

\bibitem{FuPRL2008}
\bibinfo{author}{Fu, L.} \& \bibinfo{author}{Kane, C.~L.}
\newblock \bibinfo{title}{{Superconducting Proximity Effect and Majorana
  Fermions at the Surface of a Topological Insulator}}.
\newblock \textit{\bibinfo{journal}{{Phys. Rev. Lett.}}}
  \textbf{\bibinfo{volume}{{100}}}, \bibinfo{pages}{{096407}}
  (\bibinfo{year}{{2008}}).

\bibitem{MourikScience2012}
\bibinfo{author}{Mourik, V.} \textit{et~al.}
\newblock \bibinfo{title}{{Signatures of Majorana Fermions in Hybrid
  Superconductor-Semiconductor Nanowire Devices}}.
\newblock \textit{\bibinfo{journal}{{Science}}}
  \textbf{\bibinfo{volume}{{336}}}, \bibinfo{pages}{{1003}}
  (\bibinfo{year}{{2012}}).

\bibitem{YazdaniScience2014}
\bibinfo{author}{Nadj-Perge, S.} \textit{et~al.}
\newblock \bibinfo{title}{{Observation of Majorana fermions in ferromagnetic
  atomic chains on a superconductor}}.
\newblock \textit{\bibinfo{journal}{{Science}}}
  \textbf{\bibinfo{volume}{{346}}}, \bibinfo{pages}{{602}}
  (\bibinfo{year}{{2014}}).

\bibitem{AlbrechtNature2016}
\bibinfo{author}{Albrecht, S.~M.} \textit{et~al.}
\newblock \bibinfo{title}{{Exponential protection of zero modes in Majorana
  islands}}.
\newblock \textit{\bibinfo{journal}{{Nature}}}
  \textbf{\bibinfo{volume}{{531}}}, \bibinfo{pages}{{206}}
  (\bibinfo{year}{{2016}}).

\bibitem{XuPRL2016}
\bibinfo{author}{Xu, G.}, \bibinfo{author}{Lian, B.}, \bibinfo{author}{Tang,
  P.}, \bibinfo{author}{Qi, X.-L.} \& \bibinfo{author}{Zhang, S.-C.}
\newblock \bibinfo{title}{{Topological Superconductivity on the Surface of
  Fe-Based Superconductors}}.
\newblock \textit{\bibinfo{journal}{{Phys. Rev. Lett.}}}
  \textbf{\bibinfo{volume}{{117}}}, \bibinfo{pages}{{047001}}
  (\bibinfo{year}{{2016}}).

\bibitem{PitcherJACS2010}
\bibinfo{author}{Pitcher, M.~J.} \textit{et~al.}
\newblock \bibinfo{title}{{Compositional Control of the Superconducting
  Properties of LiFeAs}}.
\newblock \textit{\bibinfo{journal}{{J. Am. Chem. Soc.}}}
  \textbf{\bibinfo{volume}{{132}}}, \bibinfo{pages}{{10467}}
  (\bibinfo{year}{{2010}}).

\bibitem{MiaoPRB2014}
\bibinfo{author}{Miao, H.} \textit{et~al.}
\newblock \bibinfo{title}{{Coexistence of orbital degeneracy lifting and
  superconductivity in iron-based superconductors}}.
\newblock \textit{\bibinfo{journal}{{Phys. Rev. B}}}
  \textbf{\bibinfo{volume}{{89}}}, \bibinfo{pages}{{220503}}
  (\bibinfo{year}{{2014}}).

\bibitem{MiaoNC2015}
\bibinfo{author}{Miao, H.} \textit{et~al.}
\newblock \bibinfo{title}{{Observation of strong electron pairing on bands
  without Fermi surfaces in LiFe$_{1-x}$Co$_x$As}}.
\newblock \textit{\bibinfo{journal}{{Nat. Commun.}}}
  \textbf{\bibinfo{volume}{{6}}}, \bibinfo{pages}{{124508}}
  (\bibinfo{year}{{2015}}).

\bibitem{WangPRB2012}
\bibinfo{author}{Wang, Z.} \textit{et~al.}
\newblock \bibinfo{title}{{Dirac semimetal and topological phase transitions in
  A$_3$Bi (A = Na, K, Rb)}}.
\newblock \textit{\bibinfo{journal}{{Phys. Rev. B}}}
  \textbf{\bibinfo{volume}{{85}}}, \bibinfo{pages}{{195320}}
  (\bibinfo{year}{{2012}}).

\bibitem{WangPRB2013}
\bibinfo{author}{Wang, Z.}, \bibinfo{author}{Weng, H.}, \bibinfo{author}{Wu,
  Q.}, \bibinfo{author}{Dai, X.} \& \bibinfo{author}{Fang, Z.}
\newblock \bibinfo{title}{{Three-dimensional Dirac semimetal and quantum
  transport in Cd$_3$As$_2$}}.
\newblock \textit{\bibinfo{journal}{{Phys. Rev. B}}}
  \textbf{\bibinfo{volume}{{88}}}, \bibinfo{pages}{{125427}}
  (\bibinfo{year}{{2013}}).

\bibitem{WangPRB2015}
\bibinfo{author}{Wang, Z.} \textit{et~al.}
\newblock \bibinfo{title}{{Topological nature of the FeSe$_{0.5}$Te$_{0.5}$
  superconductor}}.
\newblock \textit{\bibinfo{journal}{{Phys. Rev. B}}}
  \textbf{\bibinfo{volume}{{92}}}, \bibinfo{pages}{{115119}}
  (\bibinfo{year}{{2015}}).

\bibitem{WuPRB2016}
\bibinfo{author}{Wu, X.}, \bibinfo{author}{Qin, S.}, \bibinfo{author}{Liang,
  Y.}, \bibinfo{author}{Fan, H.} \& \bibinfo{author}{Hu, J.}
\newblock \bibinfo{title}{{Topological characters in Fe(Te$_{1-x}$Se$_x$) thin
  films}}.
\newblock \textit{\bibinfo{journal}{{Phys. Rev. B}}}
  \textbf{\bibinfo{volume}{{93}}}, \bibinfo{pages}{{115129}}
  (\bibinfo{year}{{2016}}).

\bibitem{DingSB2017}
\bibinfo{author}{Shi, X.} \textit{et~al.}
\newblock \bibinfo{title}{{FeTe$_{1-x}$Se$_x$ monolayer films: towards the
  realization of high-temperature connate topological superconductivity}}.
\newblock \textit{\bibinfo{journal}{{Sci. Bull.}}}
  \textbf{\bibinfo{volume}{{62}}}, \bibinfo{pages}{{503}}
  (\bibinfo{year}{{2017}}).

\bibitem{WatsonPRB2018}
\bibinfo{author}{Watson, M.~D.} \textit{et~al.}
\newblock \bibinfo{title}{{Three-dimensional electronic structure of the
  nematic and antiferromagnetic phases of NaFeAs from detwinned angle-resolved
  photoemission spectroscopy}}.
\newblock \textit{\bibinfo{journal}{{Phys. Rev. B}}}
  \textbf{\bibinfo{volume}{{97}}}, \bibinfo{pages}{{035134}}
  (\bibinfo{year}{{2018}}).

\bibitem{YajiRSI2016}
\bibinfo{author}{Yaji, K.} \textit{et~al.}
\newblock \bibinfo{title}{{High-resolution three-dimensional spin- and
  angle-resolved photoelectron spectrometer using vacuum ultraviolet laser
  light}}.
\newblock \textit{\bibinfo{journal}{{Rev. Sci. Instrum.}}}
  \textbf{\bibinfo{volume}{{87}}}, \bibinfo{pages}{{053111}}
  (\bibinfo{year}{{2016}}).

\bibitem{SatoPRL2015}
\bibinfo{author}{Kobayashi, S.} \& \bibinfo{author}{Sato, M.}
\newblock \bibinfo{title}{{Topological Superconductivity in Dirac Semimetals}}.
\newblock \textit{\bibinfo{journal}{{Phys. Rev. Lett.}}}
  \textbf{\bibinfo{volume}{{115}}}, \bibinfo{pages}{{187001}}
  (\bibinfo{year}{{2015}}).

\bibitem{SatoPRB2016}
\bibinfo{author}{Hashimoto, T.}, \bibinfo{author}{Kobayashi, S.},
  \bibinfo{author}{Tanaka, Y.} \& \bibinfo{author}{Sato, M.}
\newblock \bibinfo{title}{{Superconductivity in doped Dirac semimetals}}.
\newblock \textit{\bibinfo{journal}{{Phys. Rev. B}}}
  \textbf{\bibinfo{volume}{{94}}}, \bibinfo{pages}{{014510}}
  (\bibinfo{year}{{2016}}).

\bibitem{HughesPRL2012}
\bibinfo{author}{Chiu, C.-K.}, \bibinfo{author}{Ghaemi, P.} \&
  \bibinfo{author}{Hughes, T.~L.}
\newblock \bibinfo{title}{{Stabilization of Majorana Modes in Magnetic Vortices
  in the Superconducting Phase of Topological Insulators using Topologically
  Trivial Bands}}.
\newblock \textit{\bibinfo{journal}{{Phys. Rev. Lett.}}}
  \textbf{\bibinfo{volume}{{109}}}, \bibinfo{pages}{{237009}}
  (\bibinfo{year}{{2012}}).

\bibitem{Kresse1993}
\bibinfo{author}{Kresse, G.} \& \bibinfo{author}{Hafner, J.}
\newblock \bibinfo{title}{{Ab initio molecular dynamics for liquid metals}}.
\newblock \textit{\bibinfo{journal}{{Phys. Rev. B}}}
  \textbf{\bibinfo{volume}{47}}, \bibinfo{pages}{558} (\bibinfo{year}{1993}).

\bibitem{Kresse1996}
\bibinfo{author}{Kresse, G.} \& \bibinfo{author}{Furthm{\"u}ller, J.}
\newblock \bibinfo{title}{{Efficiency of ab-initio total energy calculations
  for metals and semiconductors using a plane-wave basis set}}.
\newblock \textit{\bibinfo{journal}{{Comput. Mater. Sci.}}}
  \textbf{\bibinfo{volume}{6}}, \bibinfo{pages}{15} (\bibinfo{year}{1996}).

\bibitem{Kresse1996prb}
\bibinfo{author}{Kresse, G.} \& \bibinfo{author}{Furthm{\"u}ller, J.}
\newblock \bibinfo{title}{{Efficient iterative schemes for ab initio
  total-energy calculations using a plane-wave basis set}}.
\newblock \textit{\bibinfo{journal}{{Phys. Rev. B}}}
  \textbf{\bibinfo{volume}{54}}, \bibinfo{pages}{11169} (\bibinfo{year}{1996}).

\bibitem{Perdew1996prl}
\bibinfo{author}{Perdew, J.~P.}, \bibinfo{author}{Burke, K.} \&
  \bibinfo{author}{Ernzerhof, M.}
\newblock \bibinfo{title}{{Generalized Gradient Approximation Made Simple}}.
\newblock \textit{\bibinfo{journal}{{Phys. Rev. Lett.}}}
  \textbf{\bibinfo{volume}{77}}, \bibinfo{pages}{3865} (\bibinfo{year}{1996}).

\bibitem{Monkhorst1976}
\bibinfo{author}{Monkhorst, H.~J.} \& \bibinfo{author}{Pack, J.~D.}
\newblock \bibinfo{title}{{Special points for Brillouin-zone integrations}}.
\newblock \textit{\bibinfo{journal}{{Phys. Rev. B}}}
  \textbf{\bibinfo{volume}{13}}, \bibinfo{pages}{5188} (\bibinfo{year}{1976}).

\end{thebibliography}
\end{document}